\documentclass[preprints,article,accept,a4paper]{mdpi}  
\pdfoutput=1

\usepackage{amssymb}
\usepackage{amsmath}
\usepackage{amsfonts}
\usepackage{amsthm}
\usepackage{graphicx}
\usepackage{color}
\newcommand{\R}{\ensuremath{\mathbb{R}}}
\graphicspath{{figures/},{./}}

\Title{Information theory for non-stationary processes with stationary increments}
\Author{Carlos Granero-Belinchon$^{(1,2)}$, St\'ephane G. Roux$^{(1)}$, Nicolas B. Garnier$^{(1)}$} 
\address{%
$^{1}$ \quad Univ Lyon, Ens de Lyon, Univ Claude Bernard, CNRS, Laboratoire de Physique, F-69342 Lyon, France; 
\\
$^{2}$ \quad ONERA-DOTA, University of Toulouse, FR-31055 Toulouse, France; 
 }%

\history{Submitted to Entropy}

\date{\today}

\abstract{
We describe how to analyze the wide class of non stationary processes with stationary centered increments using Shannon information theory.
To do so, we use a practical viewpoint and define ersatz quantities from time-averaged probability distributions. 
These ersatz versions of entropy, mutual information and entropy rate can be estimated when only a single realization of the process is available.
We abundantly illustrate our approach by analyzing Gaussian and non-Gaussian self-similar signals, as well as multi-fractal signals.
Using Gaussian signals allow us to check that our approach is robust in the sense that all quantities behave as expected from analytical derivations.
Using the stationarity (independence on the integration time) of the ersatz entropy rate, we show that this quantity is not only able to fine probe the self-similarity of the process but also offers a new way to quantify the multi-fractality.
}

\begin{document}

\maketitle 

%\tableofcontents

%%%%%%%%%%%%%%%%%%%%%%%%%%%%%%%%%%%%%%%%%%%%%%%
\section{Introduction}
\label{sec:intro}

Many real world processes, like global weather data, water reservoir levels, biological or medical signals, economic time series, {\em etc}, are intrinsicaly non-stationary~\cite{Andreas2008,Nerini2017,Boashash2013,Couts1966,Young1994}: their probability density function (PDF) deforms when time evolves. 
Analyzing such processes requires a stationary hypothesis in order to apply classical analysis, like, {\em e.g.}, two-point correlations assessment~\cite{Yang2005}. 
The stationary hypothesis can be either strict or weak: while a strict stationarity requires all moments of the process ---~and hence its PDF~--- to be time-independent, a weak stationarity is achieved when the first moment and the covariance function are time-independent and the variance is finite at all time~\cite{Debowski2007}.
Even the weaker hypothesis is often very restrictive and not realistic over long time periods. 
When the signal has a drift or a linear trend, another approach is to focus on its time-increments or time-derivatives. Indeed, assuming that the increments or time derivative are stationary is then a more realistic hypothesis.
For real world processes, the stationarity of the increments or even the stationarity of the signal is often argued to be valid when considering small chunks of data spanning short enough time range~\cite{Yaglom1955,Ibe2013,Frisch1995}, so that slow evolutions of higher order moments can be neglected.
The present article focuses on non-stationary processes with increments that are stationary and centered; this hypothesis ensures that the processes do not have any trend or drift.

Shannon information theory provides a very general framework to study stationary processes~\cite{Shannon1948,Kantz2003}, and some attempts to analyze non-stationary processes have been reported~\cite{Vu2009,Ray2010,Gmez-Herrero2015}. 
Contrary to most classical approaches, like, e.g., linear response theory in statistical physics or solid state physics, this framework is not restricted to the study of two-point correlations and linear relationships, and it allows to quantify higher order dependences~\cite{Granero-Belinchon2019} and nonlinear dynamics~\cite{Kantz2003}.
Information theory can be straightforwardly applied to any non-stationary time-process $X=\left\lbrace x_t \right\rbrace_{t\in\mathbb{R}}$: by carefully studying how the probability density and dependences of the process evolve in time, a time-evolving Shannon entropy $H_t(X)$ can be defined. The drawback of this approach is that it requires the knowledge of many realizations of the time-evolution of the process, as it relies on having enough statistics over the realizations~\cite{Gmez-Herrero2015}. 

Unfortunately, obtaining enough data is very difficult in real world systems where in the best case scenario a few realizations can be recorded experimentally, and usually only a single realization is accessible. 
In this paper, we develop a methodology that can be applied to a single realization, in order to analyse a non-stationary signal with stationary centered increments. We describe a time-averaged framework that gathers all available data points in a time window representing a single realization, whether it is the full experimental time duration, or just a fraction of it~\cite{Vu2009}.
  
The present paper is organized as follows. In section~\ref{sec:IT}, we present the general framework of information theory for a non-stationary signal, and our new framework that exploit time averages. We then give a particular emphasis on self-similar processes. In section~\ref{sec:fBm}, we report a benchmarking of our framework in the special case of Gaussian self-similar signals, a model situation where it is possible to obtain analytical developments. In section~\ref{sec:log-normal}, we explore the case of non-Gaussian self-similar processes. Finally, in section~\ref{sec:multifractal}, we drop the hypothesis of self-similarity and we apply our framework to a multifractal process.

%%%%%%%%%%%%%%%%%%%%%%%%%%%%%%%%%%%%%%%%%%%%%%%
\section{Information theory for non-stationary processes}\label{sec:ITNS}
\label{sec:IT}

%%%%%%%%%%%%%%%%%%%%%%%%%%%%%%%%%%%%%%%%%%%%%%%
\subsection{Non-stationary processes with stationary increments}\label{sec:motion}

In this article, we consider non-stationary processes with stationary increments. Such a process can be written as a motion $M=\{m_t\}_{t\in\mathbb{R}}$ obtained by integrating a stationary noise $W=\{w_t\}_{t\in\mathbb{R}}$ :
\begin{equation}
m_{t} = m_{0} + \int_{0}^{t} w_{t'} {\rm d}t' \,.
\label{eq:motion:continuous}
\end{equation}
where $m_{0}$ and $w_0$ are the values at time $t=0$, both of which can be set to 0 without loss of generality.

Nowadays signals are recorded and stored on digital media, which amounts to consider in practice a set of data sampled at discrete times $t_k$ where $k\in\mathbb{N}^+$. We further assume that the signals are equi-sampled, {\em i.e.}, $dt$ is constant and we choose $dt=1$.
So we consider in this article discrete time processes and we express them as motions $M=\{m_t\}_{t\in\mathbb{N}^+}$ obtained by integrating a stationary noise $W=\{w_t\}_{t\in\mathbb{N}^+}$ according to:
\begin{equation}
m_{t} = m_{0} + \sum_{k=1}^{t} w_{k} \,, \quad t>0 \,,
\label{eq:motion}
\end{equation}
where again $m_0=w_{0}=0$.
Eq.(\ref{eq:motion}) can also be replaced by
\begin{equation}
m_{t} = m_{t-1} + w_{t} \,, \quad t > 0 \,.
\label{eq:motion:Markov}
\end{equation}

If the noise $W$ is not centered, {\em i.e.}, has a statistical mean $\mathbb{E}(W)=\beta\neq 0$, we introduce the centered noise $w_t'=w_t-\beta$. The equations for the motion $M$ read:
\begin{align}
m_{t} &= m_0 + \beta t + \sum_{k=1}^{t} w'_{k} \,, \quad t>0 \,, \\
&= m_{t-1} + \beta + w'_{t}  \,, \quad t>0 \,.
\end{align}
The process $M$ can be interpreted as a motion built on the stationary centered noise $W'$ together with an additive deterministic drift, which is the linear trend $\beta t$.

In this article, we study motions without trend, so we impose that the noise $W$ is centered, {\em i.e.}, that its statistical mean $\mathbb{E}(W)=\beta=0$.
Besides the simple centering of the increments $W \mapsto W-\mathbb{E}(W)$, any detrending method can be applied to $M$, e.g., using moving averages.
As a consequence, the motion $M$ is centered: its statistical mean is $\mathbb{E}\{m_t\} = m_0=0$ at all times $t>0$.
Nevertheless, its variance, and all its higher order moments, may depend on time: the motion $M$ is a non-stationary process with stationary increments.
Typical examples of such processes are Brownian motion and fractional Brownian motion~\cite{Mandelbrot1968}, both of which have a variance that evolves with time.

%Processes with a deterministic trend, {\em i.e.}, of the form $m_t = \alpha + \beta t + w_t$ with $\alpha=m_0$ and $\beta=\mathbb{E}(W)$ are not motions, and are discarded, as they can be easily detrended. Detrending amounts to remove $\alpha + \beta t$, which is exactly what we assume by letting $m_0=0$ and $\mathbb{E}(W)=0$. After detrending, such a process $M$ is not a motion: it is identical to the noise $W$, which is assumed to be stationary. 

%%%%%%%%%%%%%%%%%%%%%%%%%%%%%%%%%%%%%%%%%%%%%%%
\subsection{General framework}

For a generic non-stationary process $X_t=\left\lbrace x_t \right\rbrace_{t\in\mathbb{R}}$, the probability density function (PDF) $p_{x_t}(x_t)$ changes with time.
The information theory framework can be applied to each random variable $x_t$, {\em i.e.}, at each time $t$. To do so, the PDF of $x_t$ needs to be estimated at each time $t$, which in practice requires to have many realizations available~\cite{Gmez-Herrero2015}.

To analyze the temporal dynamics of a random processes at a given time $t$, we consider the $m$-dimensional vector obtained with the Takens time-embedding procedure~\cite{Takens1981}:
\begin{equation}
\textbf{x}_t^{(m,\tau)}=\left(x_t, \,x_{t-\tau}, \, \cdots, \, x_{t-(m-1)\tau}\right)\,.
\label{eq:embed}
\end{equation}
The embedding dimension $m$ controls the order of the statistics that are considered, and the delay $\tau$ defines a time scale.
We define below some information theory quantities that are functionals of the $m$-point joint-distributions $p_{\textbf{x}^{(m,\tau)}_{t}}(\textbf{x}_t^{(m,\tau)})$, in order to characterize linear and non-linear temporal dynamics.

\medskip

\paragraph{Shannon entropy}
The entropy of $\textbf{x}_t^{(m,\tau)}$ is:

\begin{align}
H(\textbf{x}_t^{(m,\tau)}) 
&=-\int_{\mathbb{R}^{m}} p_{\textbf{x}_t^{(m,\tau)}}(\textbf{x}) \log(p_{\textbf{x}_t^{(m,\tau)}}(\textbf{x})) {\rm d}\textbf{x} \,.
\label{eq:entropy:abstract}
\end{align}

This quantity depends on time $t$, as well as on embedding parameter $m$ and delay $\tau$. We further note it $H_t^{(m,\tau)}(X) = H(\textbf{x}_t^{(m,\tau)})$, where the index $t$ indicates the time and the parameters $(m,\tau)$ are indicated as upper indices.
It measures the amount of information characterizing the $m$-dimensional PDF of the process $X$ at time $t$ sampled at scale $\tau$. When $m=1$, the entropy does not depend on $\tau$ and does not probe the dynamics of the process; we then note it $H_t(X)$, dropping the $(m=1,\tau)$ upper indices. However, for embedding dimension $m>1$ the entropy depends on the linear and non-linear dynamics of the process. Indeed, the entropy involves arbitrarily high order moments of the joint PDF $p_{\textbf{x}^{(m,\tau)}_{t}}(\textbf{x}_t^{(m,\tau)})$. As usual, the entropy  does not depend on the first moment of the distribution.

Using the time-increments of size $\tau$, $\delta_\tau x_t \equiv x_t - x_{t-\tau}$, it can be shown (see appendix \ref{sec:appendix}) that the amount of information measured by $H_t^{(m,\tau)}(X)$ is the same as the amount of information in the vector $\tilde{\textbf{x}}_t^{(m,\tau)} \equiv (x_t, \delta_\tau x_t, \delta_\tau x_{t-\tau}, ..., \delta_\tau x_{t-(m-2)\tau} )$, {\em i.e.}, 
\begin{align}
H_t^{(m,\tau)}(X) &= H(\tilde{\textbf{x}}_{t}^{(m,\tau)}) \label{eq:H:inc} \\
&=  H \left(x_t, \delta_{\tau} x_t, \delta_{\tau} x_{t-\tau}, ..., \delta_{\tau} x_{t-(m-2)\tau}\right) \,. \nonumber
\end{align}

For processes with stationary increments, the marginal distribution of $x_t$ may be strongly time-dependent, but the marginal distributions of any increment is time-independent.
Eq.(\ref{eq:H:inc}) thus suggests that the time-dependence of $H_t^{(m,\tau)}(X)$ originates mainly from $x_t$, the first component  of the rewritten embedded vector $\tilde{\textbf{x}}_t^{m,\tau}$. 
Nevertheless, it should be observed that although the $m-1$ increments, considered by themselves, have a stationary dependence structure, the covariance of $x_t$ with any of the increments is {\em a priori} non-stationary.

\paragraph{Mutual information and auto-mutual information}
The mutual information $MI$ measures the amount of information shared by two processes. 
%It quantifies by ho much information the two processus are not independent.
For two non-stationary time-embedded vectors $\textbf{x}^{(m,\tau)}_{t_1}$ and $\textbf{y}^{(n,\tau)}_{t_2}$, it is defined as :

\begin{align}
MI(\textbf{x}_{t_1}^{(m,\tau)},\textbf{y}_{t_2}^{(n,\tau)}) 
&= H_{t_1}(\textbf{x}_{t_1}^{(m,\tau)}) + H_{t_2}(\textbf{y}_{t_2}^{(n,\tau)}) \nonumber \\
&- H(\textbf{x}_{t_1}^{(m,\tau)},\textbf{y}_{t_2}^{(n,\tau)}) \,.
\end{align}

In the following, we use auto-mutual information $I_{t}^{(m,n,\tau)}(X)$ to measure, for a single process $X$, the shared information between two successive time-embedded vectors of dimension $m$ and $n$~\cite{Granero-Belinchon2017}:
\begin{align}
I_{t}^{(m,n,\tau)}(X) &= MI(\textbf{x}_t^{(n, \tau)},\textbf{x}_{t-n\tau}^{(m, \tau)}) \,.
\label{eq:AMIt}
\end{align}

\noindent Auto-mutual information defined in (\ref{eq:AMIt}) probes the dynamics of the process $X_t$ at time $t$ by measuring the dependencies between two consecutive chunks of $m$ and $n$ points sampled every $\tau$.

\medskip

\paragraph{Entropy rate}

The entropy rate, or entropy gain~\cite{Crutchfield2003}, of order $m$ at time $t$ measures the increase of Shannon entropy when the embedding dimension is increased from $m$ to $m+1$. 
It is defined as the variation of Shannon entropy between $\textbf{x}^{(m,\tau)}_{t-\tau}$ and $\textbf{x}^{(m+1,\tau)}_{t}$, two successive time-embedded versions of the process $X$:

\begin{align}
h_{t}^{(m,\tau)}(X) &= H_t^{(m+1,\tau)}(X) - H_{t-\tau}^{(m,\tau)}(X) 
\label{eq:h:abstract:diff} \\
&=  H_t(X) - I_{t}^{(m,1,\tau)}(X) \,. \label{eq:h:abstract:MI}
\end{align}

Within the general framework, the entropy, mutual information and entropy rate are well defined at any time $t$ for a non-stationary process. 
Although this framework can formally be used to analyze non-stationary processes at any time $t$, in practice it is often impossible to assess statistics at a fixed time $t$, as the number of available realizations from real world datasets may be very small.
To overcome this issue, we propose in the next section another framework that considers averages over a finite and possibly large time window, which represents for example the duration of an experimental measurement.

%%%%%%%%%%%%%%%%%%%%%%%%%%%%%%%%%%%%%%%%%%%%%%%
\subsection{Practical time-averaged framework}
\label{sec:theory_T}

We now focus on non-stationary processes with stationary increments.
We develop in this section a pragmatic approach which can be applied when a single time trace of a non-stationary signal is available.
%, {\em i.e.}, when the available dataset originates from a single trajectory.

We first present a very formal perspective that defines a time-averaged PDF of a non-stationary process.
We then propose a practical approach which uses a very simple estimation of such a time-average PDF.
We finally use this practical approach to define all the information quantities that we are interested in.

\subsubsection{Time-averaged framework}
Using a formal perspective, we consider the global statistics of the dataset, when forgetting its time dynamics, and we formally consider the time-averaged probability density function in the time window $[t_0, t_0+T]$ :
\begin{equation}
\bar{p}_{T,t_0,{\bf{x}}^{(m,\tau)}}(\textbf{x}) = \frac{1}{T} \int_{t_{0}}^{t_{0}+T} p_{{\bf x}^{(m,\tau)}_{t}}(\textbf{x}) {\rm d}t \,.
\label{eq:def:p_bar}
\end{equation}

Because of the time-average, this probability density function doesn't depend on a single time $t$ but on the starting time $t_0$ and the duration $T$ of the time window. 
%
%\begin{figure}[htb]
%\begin{center}
%\includegraphics[width=\linewidth]{pdfs_abstract}
%\caption{PDFs...
%}
%\label{fig:PDFs:theoric}
%\end{center}
%\end{figure}

In the case of a stationary process, the PDF $p_{{\bf x}^{(m,\tau)}_{t}}(\textbf{x})$ is independent of $t$, so the PDF $\bar p_{T,t_0,{\bf{x}}^{(m,\tau)}}(\textbf{x})$ is independent of $t_0$ and $T$.

In the case of a non-stationary process with stationary centered increments, the dependence on $t_0$ only appears on the mean of the time-averaged PDF $\bar p_{T,t_0,{\bf{x}}^{(m,\tau)}}(\textbf{x})$. As a consequence, since the Shannon entropy does not depend on the mean, none of the information theoretic quantities depends on $t_0$. 

In the case of a non-stationary process with stationary but non-centered increments, there is a drift: the first moment of $p_{{\bf x}^{(m,\tau)}_{t}}(\textbf{x})$ evolves linearly with time. When integrated in time in eq.(\ref{eq:def:p_bar}), this induces a deformation of the time-averaged PDF $\bar{p}_{T,t_0,{\bf{x}}^{(m,\tau)}}(\textbf{x})$, which affects {\em a priori} moments of any order. As a consequence, the Shannon entropy is then expected to depend on $t_0$. 

In the following, we focus on non-stationary processes with stationary centered increments, described in section~\ref{sec:motion}.

\subsubsection{Practical framework}

In practice, given a time series of length $T$, %we adopt a pragmatic perspective and
we propose to very roughly approximate the PDF $\bar{p}_{T,t_0,{\bf{x}}^{(m,\tau)}}$ defined in (\ref{eq:def:p_bar}) with the normalized histogram $\hat{\bar{p}}_{T,t_0}$ of all data points $x_t^{(m,\tau)}$, $t \in [t_0, t_0+T]$ available in the time window. 
This is a very strong assumption, as $\bar{p}_{T,t_0,{\bf{x}}^{(m,\tau)}}$ is {\em a priori} very different from any $p_{x_t^{(m,\tau)}}$, and {\em a priori} very different from the histogram $\hat{\bar{p}}_{T,t_0}$ constructed after cumulating all the available data in the interval. 
This pragmatic approach comes down to treat the set of available data points in the time interval exactly in the same way as if it was a set of data points originating from a stationary, albeit unknown, process and then estimate its PDF. 

In the following, we drop the hat in the notations, and consider only the ersatz probabilities $\hat{\bar{p}}$ in place of the time-averaged probabilities ${\bar{p}}$. 
As we discuss later in section~\ref{sec:discussion}, if several experimental realizations are available, it is of course possible to use them to enhance the estimation of the time-averaged PDF.

\subsubsection{Information theory quantities in the practical framework}

Given a time series of length $T$, and considering the ersatz PDFs $\bar{p}_{T,t_0,{\bf{x}}^{(m,\tau)}}$, we define $\bar{H}_{T}^{(m,\tau)}(X)$ the time-averaged Shannon entropy, $\bar I_{T}^{(m,n,\tau)}(X)$ the time-averaged auto-mutual information and $\bar h_{T}^{(m,\tau)(X)}$ the time-averaged entropy rate, as described below. 

\paragraph{Ersatz Shannon entropy}
We define the ersatz entropy of the time-embedded signal as the entropy of the time-averaged PDF $\bar{p}_{T,t_0,\textbf{x}^{(m,\tau)}}$:

\begin{equation}
\bar{H}_{T}^{(m,\tau)}(X)=-\int_{\mathbb{R}^{m}} \bar p_{T,t_0,\textbf{x}^{(m,\tau)}}(\textbf{x}) \log(\bar p_{T,t_0,\textbf{x}^{(m,\tau)}}(\textbf{x})) {\rm d}\textbf{x} \,.
\label{eq:def:entropy:ersatz}
\end{equation}

$\bar{H}_{T}^{(m,\tau)}(X)$ gives the amount of information of the set of values of the signal ${\textbf x}_t^{(m,\tau)}$ in the time interval $[t_0, t_0+T]$ and hence it can be interpreted as the total information characterizing the temporal trajectory $\{\textbf{x}^{(m,\tau)}_t, t \in [t_0, t_0+T]\}$ of the process. 
If the process has stationary centered increments, the total amount of information in the trajectory depends only on its length $T$, and not on its starting time $t_0$.
In that sense, the ersatz entropy $\bar{H}_{T}^{(m,\tau)}$ is not stationary. 

Using the rewriting (\ref{eq:H:inc}), we argue that this dependence in $T$ originates from $x_t$ ---~the first component of the vector $\tilde{\textbf{x}}_t^{m,\tau}$~--- which has a time-dependent marginal distribution. 
Because the $m-1$ other components of $\tilde{\textbf{x}}_t^{m,\tau}$ are increments, they have by hypothesis a stationary dependence structure. So increasing the embedding dimension does not impact the dependence of the ersatz entropy on the window size $T$, but only its dependence on the increments size $\tau$.

\paragraph{Auto-mutual information}
We define the ersatz auto-mutual information as:
\begin{equation}
\bar I_{T}^{(m,n,\tau)}(X)=  \bar{H}_{T}^{(m,\tau)}(X) + \bar{H}_{T}^{(n,\tau)}(X) 
- \bar{H}_{T}^{(m+n,\tau)}(X) \,.
\label{eq:def:AMI:ersatz}
\end{equation}

\paragraph{Entropy rate}

We define the ersatz entropy rate over a time interval of size $T$ as:
\begin{align}
\bar h_{T}^{(m,\tau)}(X) 
&= \bar{H}^{(m+1,\tau)}_{T}(X) - \bar{H}^{(m,\tau)}_{T}(X) 		\label{eq:def:h:diff:ersatz} \\
&=  \bar{H}_{T}(X) - \bar I_{T}^{(m,1,\tau)}(X) \,. 			\label{eq:def:h:ersatz}
\end{align}

From (\ref{eq:def:h:diff:ersatz}), we may expect a cancelation of the main dependence in $T$ which is the same for $\bar{H}_{T}^{(m+1,\tau)}$ and for $\bar{H}_{T}^{(m,\tau)}$. As a consequence, the ersatz entropy rate $\bar{h}_{T}^{(m,\tau)}$ should be stationary, in the sense that it should not depend on the length $T$ of the time interval that is considered.

If the available samples span a very large time window, one may consider using multiple non-overlapping time windows of size $T$ starting at various times. Because of the stationarity and zero-mean of the increments and hence the independence of the ersatz quantities on $t_0$, it is possible to average the different estimations of the ersatz quantities obtained in each window. It is also possible to use all the non-overlapping windows to populate the histogram and thus enhance the estimation of the time-averaged PDF. Each of these two operations will increase the statistics and hence improve the estimation.

%%%%%%%%%%%%%%%%%%%%%%%%%%%%%%%%%%%%%%%%%%%%%%%
\subsection{Self-similar processes}

In this section, we focus on the special case of self similar processes, {\em i.e.}, signals which exhibit monofractal scale invariance~\cite{Mandelbrot1982}. 
Such processes have been used as a satisfying first approximation to model or describe very various phenomena, such as ionic transport~\cite{Mauritz1989}, fluid turbulence~\cite{Chevillard2012}, climate~\cite{Kavvas2015}, river flows~\cite{Rigon1996}, cloud structure~\cite{Gotoh1998} or earthquakes~\cite{Console2003},
as well as neural signals~\cite{Ivanov2009}, stock markets~\cite{Drozdz1999,Cont1997}, texture patterns~\cite{Uhl2015} or internet traffic~\cite{Chakraborty2004}.
A process $X_t$ is monofractal scale-invariant if there exists a real number $\mathcal{H}$ such that for all $a\in \R^{+*}$, the probability density functions of $x_{at}$ and $a^{\mathcal{H}} x_t$ are equivalent.  $\mathcal{H}$ is called the Hurst exponent. 
If ${\cal H}<0$, the process is stationary and called a fractional noise.
If $0\le {\cal H}<1$, the process is non-stationary with stationary increments. 
The case ${\cal H}=1/2$ corresponds to the traditional Brownian motion.

Assuming $x_{t=0}=0$, the scale invariance property can be expressed as~\cite{Flandrin1992}:
\begin{equation}
p_{x_{at}}(x)=\frac{1}{a^{\mathcal{H}}} p_{x_t}\left(\frac{x}{a^{\mathcal{H}}}\right) \,.
\end{equation}
The scale invariance property of a process $X_t$ transfers to its increments, as well as any of its time-embedded version: 
\begin{equation}
p_{\textbf{x}^{(m,a\tau)}_{at}}({\bf x})=\frac{1}{a^{m\mathcal{H}}} p_{\textbf{x}^{(m,\tau)}_{t}} \left(\frac{\bf x}{a^{\mathcal{H}}}\right).
\label{eq:self_similar}
\end{equation}

This relation allows to express the non-stationary PDF of $\textbf{x}_t^{(m,\tau)}$ at any time $t$ as a function of the PDF at unit-time ($t$=1). This is done by using the factor $a=1/t$ in eq.(\ref{eq:self_similar}), {\em i.e.}, by rescaling each coordinate of the embedded vector by the factor $t^{\cal H}$. 

Using eq.(\ref{eq:def:p_bar}), it is straightforward to see that the scale invariant property of the form (\ref{eq:self_similar}) is also valid for the time-averaged PDF $\bar{p}_{T,t_0,{\bf{x}}^{(m,\tau)}}(\textbf{x})$.

Because of its definition (\ref{eq:motion}) as a cumulative sum of a noise, a motion can be seen as accumulating the correlations between successive points of the noise.
When performing a time-embedding, the particular case $\tau=1$ is interesting: considering the relation (\ref{eq:H:inc}), we may expect that the information contained in the time-embedded motion $\textbf{m}_t^{(m,\tau=1)}$ is closely related to the information contained in the time-embedded noise $\textbf{w}_t^{(m,\tau=1)}$. This is not the case anymore when $\tau \ge 2$.

%%%%%%%%%%%%%%%%%%%%%
\paragraph*{Fractional Brownian motion}

The Fractional Brownian motion (fBm) was proposed by Mandelbrot and Van Ness~\cite{Mandelbrot1968} and quickly became a benchmark for self-similarity and long-range dependence. 
The fBm is the only Gaussian self-similar process with stationary increments. It is characterized by its Hurst exponent, $\mathcal{H}$.

The fBm is a motion, obtained by integrating according to (\ref{eq:motion}) a fractional Gaussian noise (fGn), defined as a centered Gaussian process with the correlation structure
\begin{align}
 c_{\rm fGn}(\tau)&=\frac{\sigma_1^2}{2} \left[(\tau-1)^{2{\cal H}}-2\tau^{2{\cal H}}+(\tau+1)^{2{\cal H}}\right] \,.
 \label{eq:fGn:correlation}
\end{align}
The fGn is a stationary noise with the standard deviation $\sigma_1$. It is scale-invariant with a Hurst exponent ${\cal H}-1$.

The non-stationary covariance structure of the fBm $B$ reads 
\begin{equation}\label{eq:fbmcorr}
\mathbb{E} \{B_t B_{t-\tau} \} = \frac{\sigma_1^2}{2} \left[t^{2\mathcal{H}}+(t-\tau)^{2\mathcal{H}}-\tau^{2\mathcal{H}}\right]   \,,
\end{equation}
where $\tau<t$.

%%%%%%%%%%%%%%%%%%
\subsubsection{General framework}

We show below how the theoretical information quantities depend on time $t$ and delay $\tau$.
We start from the relation (\ref{eq:H:inc}) between the entropy of the time-embedded vector and the entropy of the increments
and we normalize each component of the vector $\tilde{\textbf{x}}_t^{(m,\tau)}$ by its standard deviation. 
The standard deviation $\sigma_t$ of the motion $x_t$ evolves with time as $\sigma_t = \sigma_1 t^{\cal H}$, while the standard deviation $\sigma_\tau$ of the increments $x_t-x_{t-\tau}$ is independent of $t$, thanks to the stationarity of the increments, and evolves with the size of the increment as $\sigma_\tau = \sigma_1 \tau^{\cal H}$. So we have:
\begin{align}
H_{t}^{(m,\tau)}(X) 
&=  H \left(x_t/\sigma_t, \delta_{\tau} x_t / \sigma_\tau, ..., \delta_{\tau} x_{t-(m-2)\tau}/\sigma_\tau\right) 
\nonumber \\
 &+ \ln \sigma_t + (m-1)\ln \sigma_\tau  \,, \\
&=  H \left(x_t/t^{\cal H}, \delta_\tau x_t / \tau^{\cal H}, ..., \delta_{\tau} x_{t-(m-2)\tau}/\tau^{\cal H}\right) 
\nonumber \\
 &+ {\cal H}\ln t + (m-1){\cal H}\ln \tau  + m \ln \sigma_1\,.
\end{align}
We then use the scaling law (\ref{eq:self_similar}) for $a=1/t$ to relate the joint probability at a given time $t$ to the joint probability at unit-time $t=1$, which leads to:
\begin{align}
H \left(x_t/t^{\cal H}, \delta_\tau x_t / \tau^{\cal H}, ..., \delta_{\tau} x_{t-(m-2)\tau}/\tau^{\cal H}\right) 
%\nonumber \\
&=H \left(x_1, \delta_{\tau/t} x_1 , ..., \delta_{\tau/t} x_{1-(m-2)\tau/t}\right) 
- m\ln \sigma_1 \\
&=H\left(\tilde{\textbf{x}}_{t=1}^{(m,\tau/t)}\right)
- m\ln \sigma_1 \,.
\end{align}
Using (\ref{eq:H:inc}) again at time $t=1$, we have $H\left(\tilde{\textbf{x}}_{t=1}^{(m,\tau/t)}\right) = H_{t=1}^{(m,\tau/t)}(X)$,
so we can express the time-dependent Shannon entropy (\ref{eq:entropy:abstract}) for self-similar processes as:
\begin{align}
H_t^{(m,\tau)}(X) &= H_{t=1}^{(m,\tau/t)}(X) + {\cal H}\ln t + (m-1){\cal H}\ln \tau  \,.
\label{eq:entropy:abstract:ssimilar}
\end{align}

The entropy rate can be rewritten with (\ref{eq:h:abstract:diff}) and (\ref{eq:entropy:abstract:ssimilar}) as:
\begin{equation}
h_t^{(m,\tau)}(X) = h_{t=1}^{(m,\tau/t)}(X) + {\cal H}\ln \tau \,,
\label{eq:h:abstract:ssimilar}
\end{equation}
where $h_1^{(m,\tau/t)}(X)$ is the entropy rate at time $t=1$, using the rescaled time delay $\tau/t$. 

Although the two quantities $H_{t=1}^{(m,\tau/t)}(X)$ and $h_{t=1}^{(m,\tau/t)}(X)$ are considered at a fixed time $t=1$, they still depend on $t$ via the delay $\tau/t$. Because $\tau/t$ is small as soon as $t\gg\tau$, we expect that the dependence of the entropy $H_t^{(m,\tau)}(X)$ on time $t$ is mainly in ${\cal H}\ln t$, and that the entropy rate is almost time-independent.

%%%%%%%%%%%%%%%%%%%%%
\paragraph*{Fractional Brownian motion}

The PDF $p_{B_t}$ of the fBm is Gaussian at any time $t$, so we can express its Shannon entropy and entropy rate at time $t$ by using eq.(\ref{eq:entropy:abstract:ssimilar}) and the expression of the Shannon entropy of a Gaussian multivariate process~\cite{Zografos2005}. We obtain the following approximated expressions:
\begin{align}
H_t^{(m,\tau)}(B) &\simeq m H_1^{\rm FBM} + {\cal H}\ln t + (m-1){\cal H}\ln \tau 
\label{eq:H:fBm}
\\  
h_t^{(m,\tau)}(B) &\simeq H_1^{\rm FBM} + {\cal H}\ln \tau \,,
\label{eq:h:fBm}
\end{align}
% XII.255
%
where $H_1^{\rm FBM}\equiv \frac{1}{2}\ln \left( 2\pi e \sigma_1^2\right)$ is the entropy of the fBm at unit-time.
These formulae are exact for $m=1,2$, but for $m\ge 3$, constant terms as well as corrections in $\tau/t$ have been omitted for clarity. 

%%%%%%%%%%%%%%%%%%%%%
\subsubsection{Practical time-averaged framework} 

For a generic self-similar process, we are not able to derive any analytical results in the practical time-averaged framework. 
Nevertheless, the behaviors expected for a generic non-stationary process with stationary increments are holding: 
i) the ersatz entropy $\bar{H}_{T}^{(m,\tau)}$ is not stationary, in the sense that it depends on the length $T$ of the time-interval,
ii) the ersatz entropy rate $\bar{h}_{T}^{(m,\tau)}$ is stationary.

%%%%%%%%%%%%%%%%%%%%%
\paragraph*{Fractional Brownian motion}

The ersatz entropy of the fBm over a time window of size $T \gg \tau$ can be expressed by averaging its covariance structure on a time window of size $T$. We obtain~\cite{GBelinchon2016}:
\begin{equation}
\bar{H}_{T}^{(1,\tau)}(B) = H_1^{\rm FBM} + \mathcal{H} \ln T \,.
\label{eq:H_T:fBm}
\end{equation}
The entropy of the fBm thus increases linearly with the logarithm of the window size $T$. The larger the time window, the more there is information in the trajectory. 

The auto-mutual information of the fBm can be derived in the same way using (\ref{eq:def:AMI:ersatz}) for $T\gg\tau$: 
\begin{equation}
\bar{I}_{T}^{(1,1,\tau)}(B)= - \mathcal{H} \ln \left(\frac{\tau}{T}\right) + {\cal C}\left( \frac{\tau}{T}\right) \,,
\label{eq:AMI_T:fBm}
\end{equation}
where ${\cal C}\left( \frac{\tau}{T}\right)$ is a correction in $\tau/T$ that reads
\begin{align}
{\cal C}\left( \frac{\tau}{T}\right)
&= \frac{1}{2}\ln\left(\frac{\frac{\tau}{2T}+\frac{1}{2\mathcal{H}+1}}{(\frac{\tau}{T}+\frac{1}{2\mathcal{H}+1})}\right)
\label{eq:correction:full}  \\
&= - \frac{2{\cal H}+1}{4}\frac{\tau}{T} + {\cal O}\left( \left(\frac{\tau}{T}\right)^2\right)\,.
\label{eq:correction:dev}
\end{align}
The ersatz mutual information depends logarithmically on the scale $\tau$ and the window size $T$. The larger the window-size $T$ or the smaller the scale $\tau$, the stronger the dependencies.

The ersatz entropy rate of order $m=1$ is obtained by combining (\ref{eq:H_T:fBm}) and (\ref{eq:AMI_T:fBm}) according to (\ref{eq:def:h:ersatz}):
\begin{equation}
\bar{h}_{T}^{(1,\tau)}(B) =  H_1^{\rm FBM} + \mathcal{H} \ln \tau - {\cal C}\left( \frac{\tau}{T}\right)
\label{eq:h_T:fBm}
\end{equation}
which is independent of $T$ up to corrections in $\tau/T$, while being linear in $\ln(\tau)$ with a constant slope ${\cal H}$.
The correction $- {\cal C}({\tau}/{T})$ in eq.(\ref{eq:h_T:fBm}) is positive, see eq.(\ref{eq:correction:dev}).

Comparing (\ref{eq:H:fBm}) with (\ref{eq:H_T:fBm}) shows that for the fBm, the ersatz entropy dependence on $T$ is exactly the same as the entropy dependence on $t$. 
Comparing (\ref{eq:h:fBm}) with (\ref{eq:h_T:fBm}) shows that the entropy rate and the ersatz entropy rate do not depend on $t$ or $T$ up to corrective terms that are negligible if the scale $\tau$ is not too large. 
We also see explicitly that both quantities evolves with the scale $\tau$ in $\mathcal{H} \ln \tau$, again up to corrections of order $\tau/t$ and $\tau/T$.

The example of the fBm suggests that for a scale-invariant process the evolution of any information theory quantity with the scale $\tau$ is the same within the practical time-averaged framework or the general framework.
We push this analysis further in the next sections, by exploring if this property holds when the process is non-Gaussian.

%%%%%%%%%%%%%%%%%%%%%%%%%%%%%%%%%%%%%%%%%%%%%%%%%%%%%%%%%%%%
%%%%%%%%%%%%%%%%%%%%%%%%%%%%%%%%%%%%%%%%%%%%%%%%%
\section{Benchmarking the practical framework with the fBm}
\label{sec:fBm}

We focus in this section on the fractional Brownian motion, for which analytical expressions were derived in the previous sections.
We use the fBm not only to benchmark our estimators of information theory quantities, but also to illustrate the use of the practical framework and the expected behavior of the ersatz quantities when used on a self-similar process of Hurst exponent ${\cal H}$.

\subsection{Characterization of the estimates}
\label{sec:characterization}

\subsubsection{Data}
\label{sec:data}

To obtain a fBm, we integrate a fractional Gaussian noise (fGn). We use circulant matrix method~\cite{Helgason2011} to impose the correlation structure of the fGn (\ref{eq:fGn:correlation}) .
Then, we center and normalize the noise such that the standard deviation, $\sigma_{\rm fGn}$, is equal to one. We then take the cumulative sum to obtain the fBm. 
Through all this article, ${\cal H}=0.7$ for all the processes used to illustrate our results, but we have checked that they hold for any other value $0<{\cal H}<1$.

%One can interpret the time lag $\tau$ as defining a coarse-grained time scale at which the signal is down-sampled. This approach has proven fruitful in interpreting the scaling of the entropy rate of the velocity field in Turbulence data~\cite{Granero16}, a famous real-world example of long range motion process.

\subsubsection{Procedure}

We estimate the Shannon entropy $\bar{H}_{T}^{(m,\tau)}$ with our own implementation of the $k$-nearest neighbors estimate from Kozachenko and Leonenko~\cite{L.Kozachenko1987}.
We estimate the auto-mutual information $\bar{I}_{T}^{(m,p,\tau)}$ with the algorithm provided by Kraskov, Stogbauer and Grassberger~\cite{Grassberger2004}. This estimator is also based on a nearest neighbors search and it provides ---~amongst several good properties~--- a build-in cancellation of the bias difference originating from each of the two arguments.
In the following, we note $k$ the number of neighbors, which is the only parameter of the estimators.
The entropy rate $\bar{h}_{T}^{(m,\tau)}$ is then computed using eq.(\ref{eq:def:h:ersatz}). %, unless noted otherwise.

We generate for each motion a set of $100$ independent realizations of fixed size $T$ with a Hurst exponent ${\cal H}=0.7$. 
%We use $N=2^{16}$ points. 
We compute averages of the estimates on the realizations and use the standard deviation as error bars in the different graphs.

In subsections \ref{sec:subsec:bias} and \ref{sec:subsec:std}, we characterize respectively the bias and standard deviation (std) over realizations of our estimators of entropy, auto-mutual information and entropy rate.

\subsubsection{Convergence / bias}\label{sec:subsec:bias}

\begin{figure}[htb]
\begin{center}
\includegraphics[width=\columnwidth]{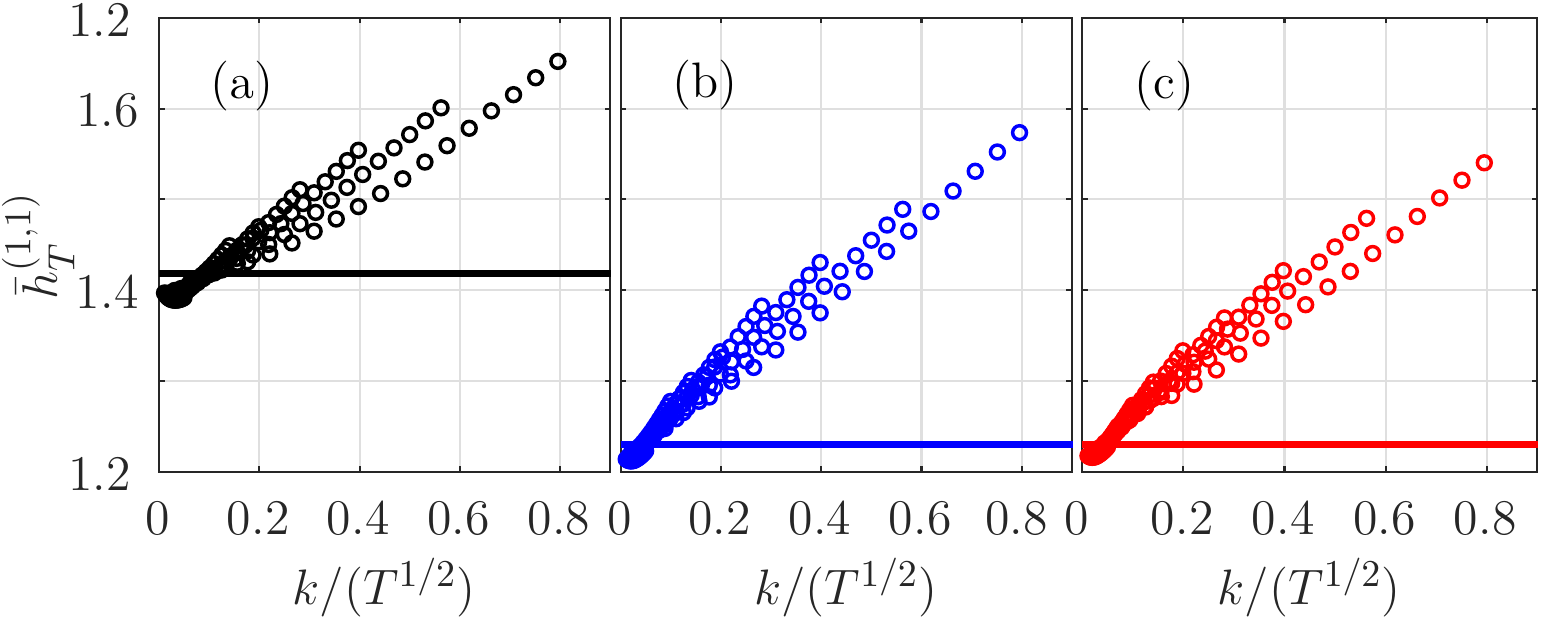}
\caption{
Dependence of $\bar{h}_{T}^{(1,\tau)}$ (for $\tau=1)$) on $k/T^{\frac{1}{2}}$ for the fBm (a, in black) and for the Hermitian (b, in blue) and even-Hermitian (c, in red) log-normal processes.}
\label{fig:bias}
\end{center}
\end{figure}

We detail here how the ersatz entropy rate evolves with $T$ and $k$. We report in Fig.~\ref{fig:bias}a our results for all possible values of the couples $(\log_2(T),k) \in [9, ..., 17]\times[4, ..., 18]$, while $\tau$ is set to 1 here.
According to eq.(\ref{eq:h_T:fBm}), the ersatz entropy rate of the fBm converges for large $T$ to the value $H^{\rm fBm}_1$ (horizontal black line in Fig.~\ref{fig:bias}a) thanks to the vanishing of the correction term ${\cal C}(\tau/T)$, according to (\ref{eq:correction:dev}).

Fig.~\ref{fig:bias}a can be interpreted as describing the behavior of the bias of the estimator.
This bias vanishes non-monotonically as $k/T^\frac{1}{m+1}$.
When $k/T^\frac{1}{m+1}$ is reduced, first the bias is positive and diminishes toward negative values and then converges to zero.
This behavior was previously reported for the $k$-nn mutual information estimator applied for stationary processes~\cite{Grassberger2004,Gao2016,Granero-Belinchon2019}, and we confirm it is valid for the fBm. 

We observed the same convergence for a large range of scales $\tau>1$: the ersatz entropy rate then converges to $H^1_{\rm fBm} + {\cal H}\ln\tau$ for large $T$ with the same behavior of the bias.

\subsubsection{Standard deviation of the estimates}\label{sec:subsec:std}

\begin{figure}[htb]
\begin{center}
\includegraphics[width=\linewidth]{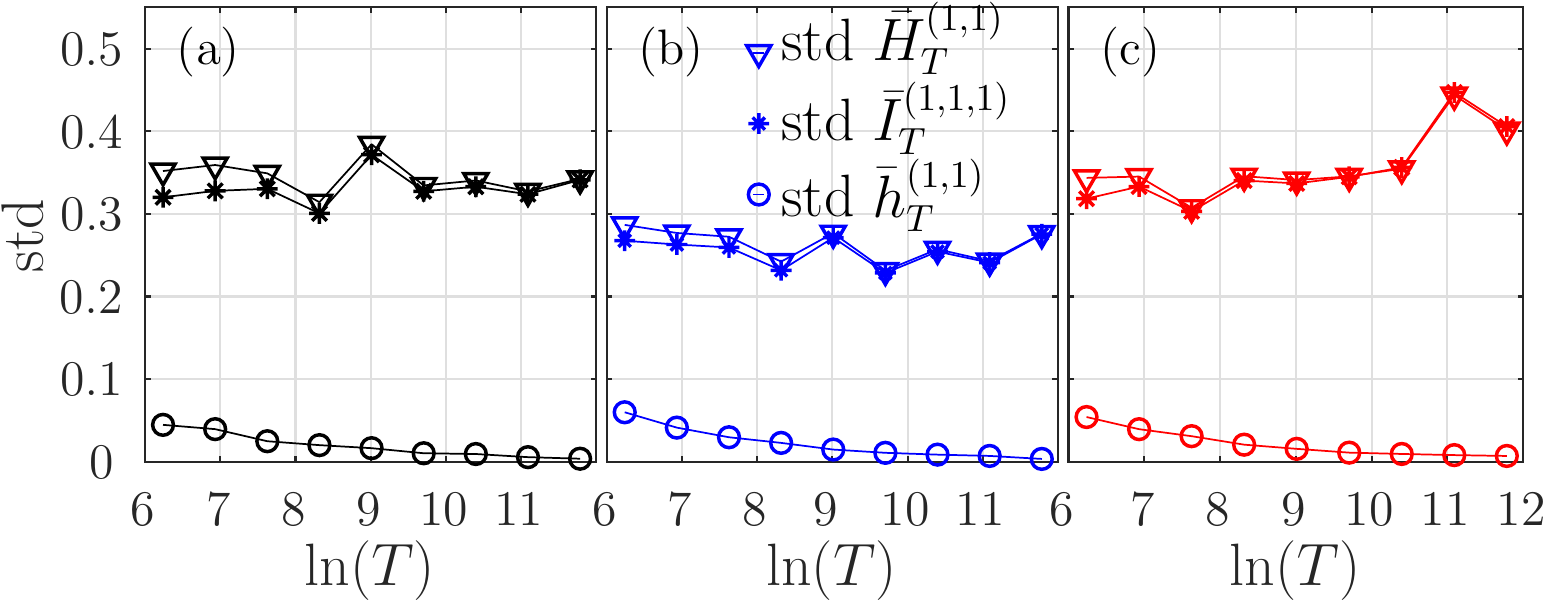}
\caption{Standard deviations of $\bar{H}_{T}^{(1,\tau)}$ (triangles), $\bar{I}_{T}^{(1,1,\tau)}$ (circles) and $\bar{h}_{T}^{(1,\tau)}$ (stars), for $\tau=1$, as functions of $T$,
for the fBm (a, black) 
the Hermitian (b, in blue) and even-Hermitian (c, in red) log-normal processes.
}
\label{fig:std}
\end{center}
\end{figure}

We present in Fig.~\ref{fig:std}a the evolution of the standard deviation of the ersatz entropy, mutual information and entropy rate with $T$ for $\tau=1$. The standard deviation of both the entropy and mutual information is large, and does not decrease when $T$ ---~and hence the number of samples~--- increases. 
On the contrary, the standard deviation of the entropy rate is much smaller and decreases when $T$ increases.
We attribute this feature to the dependence of the quantities on the observation time $T$, see eqs.(\ref{eq:H_T:fBm}) and (\ref{eq:AMI_T:fBm}) for the fBm. While $\bar{H}_T$ and $\bar{I}_T$ increase as $\ln T$, this is not the case for $\bar{h}_T$ which is independent on $T$ (up to small corrections, negligible for smallish $\tau$).
Although it is difficult to explain why the standard deviation of the entropy and mutual information remain constant when $T$ increases, it seems that this results from a balance between the non-stationarity (in $\ln T$) and the increased statistics. On the contrary, for the entropy rate which is stationary, the decrease of the std is as expected.

\medskip

As a conclusion, both the bias and the standard deviation of the ersatz entropy rate increase when $k$ increases or $T$ decreases and can be made arbitrarily small by increasing the window size $T$. 
In the remainder of this article, we choose $k=5$ and when studying the behavior of information theoretic quantities on the scale $\tau$, we set $T=2^{16}$.

%%%%%%%%%%%%%%%%%%%%%%%%%%%%%%%%%%%%%%
\subsection{Dependence on times $T$ and $\tau$}

In this section, we present a detailed numerical study of the ersatz entropy, auto-mutual information and entropy rate of the fBm with ${\cal H}=0.7$. In particular, we present a quantitative comparison with the analytical expressions (\ref{eq:H:fBm}, \ref{eq:h:fBm}) in the general framework, as well as with analytical expressions~(\ref{eq:H_T:fBm}, \ref{eq:AMI_T:fBm}, \ref{eq:h_T:fBm}) in the practical framework for the fBm. 
These comparisons allow: first, to validate the analytical expressions obtained for fBm in the practical framework, and second to show that the information theoretic quantities in the practical framework evolve in $T$ and $\tau$ exactly as their counterparts evolve in the general framework in $t$ and $\tau$. 
To compare analytical and numerical results, we vary the window size $T$, the scale $\tau$ and the embedding dimension $m$. 

\subsubsection{Entropy and auto-mutual information}

\begin{figure}
\begin{center}
\includegraphics[width=\linewidth]{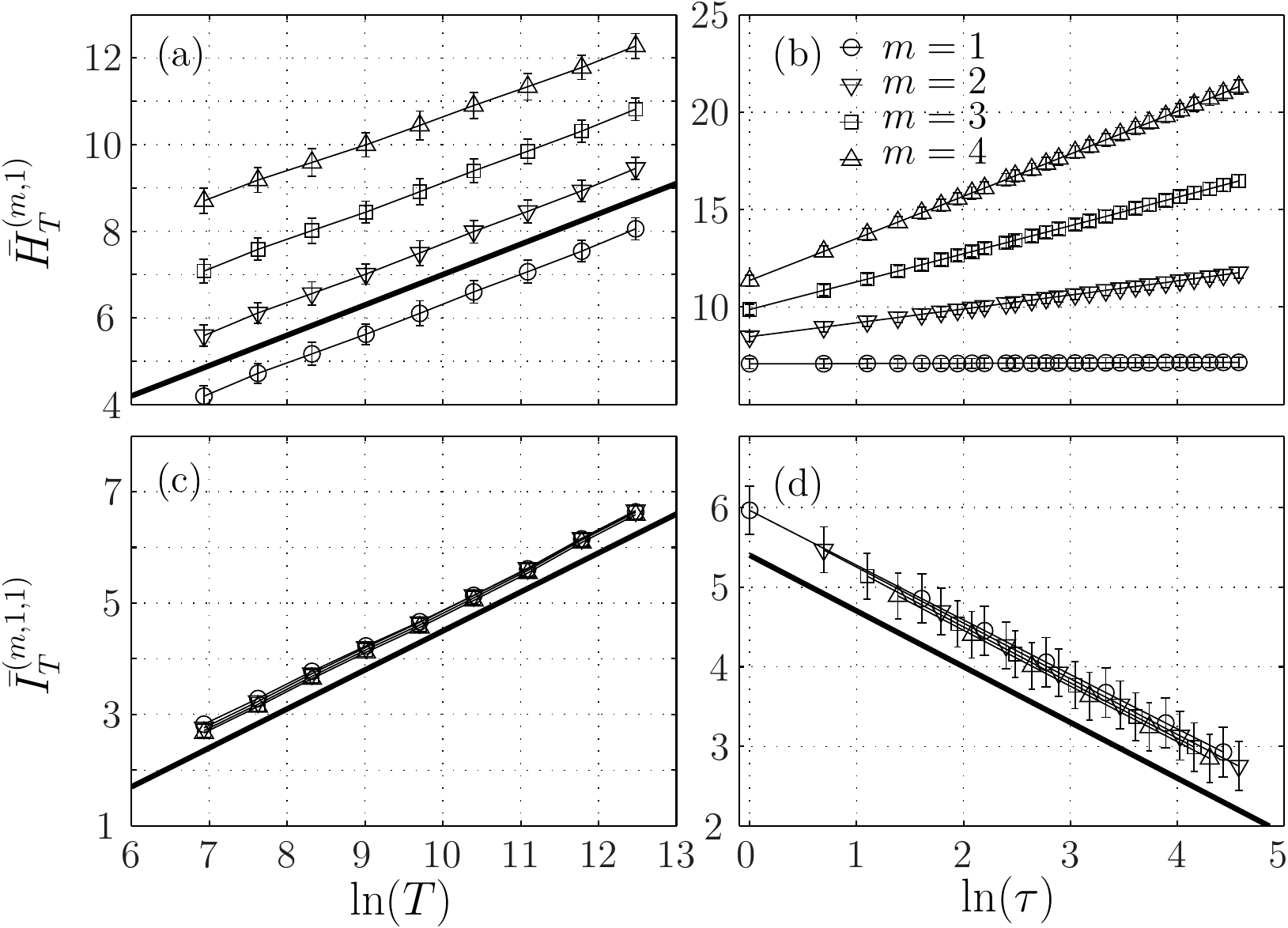}
\caption{(a) Entropy $\bar{H}_{T}^{(m,\tau)}$ and (c) auto-mutual information $\bar{I}_{T}^{(m,1,\tau)}$ of the fBm in function of the logarithm of the window size $\ln(T)$ for a fixed scale $\tau=1$. (b) Entropy  and (d) auto-mutual information in function of the logarithm of the scale of analysis $\ln(\tau)$ for a fixed $T=2^{16}$.  
Each symbol corresponds to a different embedding dimension $m$.
In (a) and (c) the black line has a slope $\mathcal{H}=0.7$, while in (d) its slope is $-\mathcal{H}=-0.7$.
}
\label{fig:T:tau:fBm}
\end{center}
\end{figure}

\begin{figure}
\begin{center}
\includegraphics[width=\linewidth]{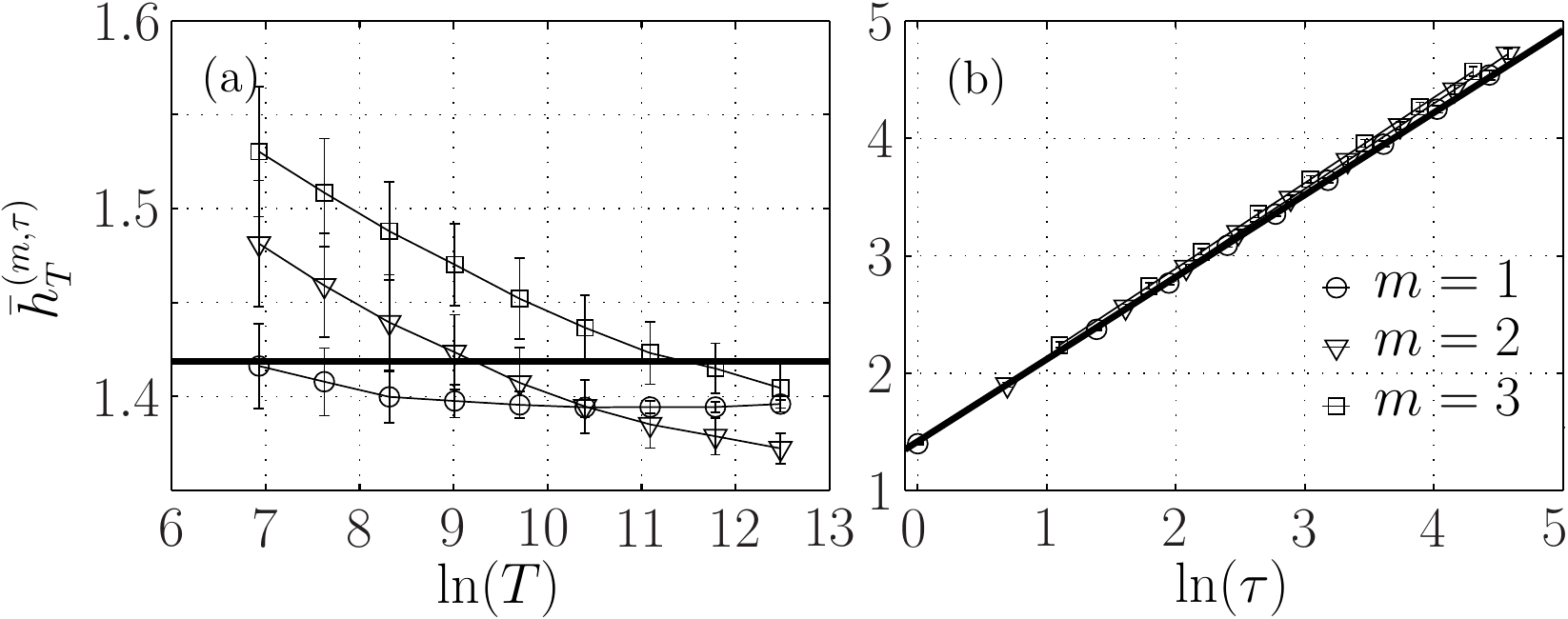}
\caption{Ersatz entropy rate $\bar{h}_{T}^{(m,\tau)}$ of a fBm with $\mathcal{H}=0.7$.
(a): as a function of the window size $T$ for fixed $\tau=1$.
(b) : as a function of the scale $\tau$ for $T=2^{16}$. 
Each symbol corresponds to a different embedding dimension $m$.
The horizontal black line in (a) indicates the theoretical value $H_1^{\rm fBm}$.
The black line in (b) represents the linear function $H_1^{\rm fBm} + \mathcal{H}\ln\tau$ with $\mathcal{H}=0.7$.
}
\label{fig:h:tau:fBm}
\end{center}
\end{figure}

\paragraph{Dependence on $T$}
The left column of Fig.~\ref{fig:T:tau:fBm} shows the ersatz Shannon entropy $\bar{H}_{T}^{(m,\tau)}$ (Fig.~\ref{fig:T:tau:fBm}a) and auto-mutual information $\bar{I}_{T}^{(m,1,\tau)}$ (Fig.~\ref{fig:T:tau:fBm}c) at a given scale $\tau=1$, as a function of $\ln T$. 
The evolution of these two quantities for $m=1$ is very close to $\mathcal{H}\ln T$, which is represented by a continuous black line. This is in agreement with eq.(\ref{eq:H_T:fBm}) and eq.(\ref{eq:AMI_T:fBm}). For $m>1$, we obtain in the practical framework the behaviors predicted in the general framework, replacing $t$ by $T$ in the equations.
We observe that the auto-mutual information does not depend on the embedding dimension $m$, while the entropy does, with an offset that seems to depend linearly on $m$.
The dependence of the entropy and the auto-mutual information on the time window $T$ is the signature of the non-stationarity of the signal.

\paragraph{Dependence on $\tau$}
The right column of Fig.~\ref{fig:T:tau:fBm} shows the ersatz Shannon entropy and auto-mutual information for a fixed window size $T=2^{16}$ when varying the scale parameter $\tau$.
The ersatz Shannon entropy behaves as $(m-1)\mathcal{H}\ln(\tau)$, see Fig.~\ref{fig:T:tau:fBm}b, in agreement with eq.(\ref{eq:entropy:abstract:ssimilar}) or eq.(\ref{eq:H:fBm}).
The ersatz auto-mutual information behaves as $-\mathcal{H}\ln(\tau)$ for any embedding $m$, see Fig.~\ref{fig:T:tau:fBm}d, in agreement with eq.(\ref{eq:AMI_T:fBm}), thus suggesting this formula is valid for any embedding dimension.

\subsubsection{Stationarity of the entropy rate}

\begin{figure}
\begin{center}
\includegraphics[width=\linewidth]{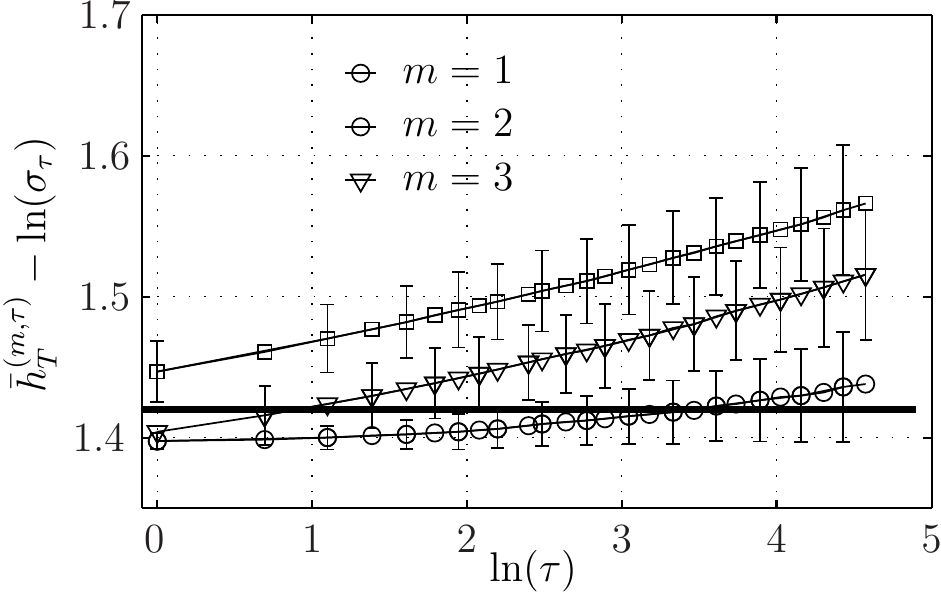}
\caption{Ersatz entropy rate $\bar{h}_{T}^{(m,\tau)}-\ln(\sigma_{\tau})$ of the fBm as a function of $\ln(\tau)$ for fixed $T=2^{16}$ and varying embedding dimension $m$.
The thick horizontal black line represents the constant value $H_1^{\rm fBm}$.
}
\label{fig:h-hurst}
\end{center}
\end{figure}

Fig.~\ref{fig:h:tau:fBm}a shows that the ersatz entropy rate $\bar{h}_{T}^{(m=1,\tau)}$ with embedding dimension $m=1$ is almost constant when $T$ is varied. For embedding dimensions $m>1$, there is a small variation, of about 15$\%$, much smaller than the 200\% variation observed for either the entropy or the auto-mutual information (Fig.~\ref{fig:T:tau:fBm}a,c) on the same range of $T$.
This small dependence on $T$ can be due to the correction in eq.(\ref{eq:h_T:fBm}), which may depend on $m$. 
We argue that it is mostly due to bias, which increases with the embedding dimension.
Indeed, we observe that the entropy rate seems to converge for larger $T$ to the same value close to $H_1^{\rm fBm}$ for all $m$. 
As a larger $T$ corresponds to a larger sampling of the statistics, the bias is reduced, as reported in Fig.~\ref{fig:bias}. 
Moreover, for $m=1$, eq.(\ref{eq:h_T:fBm}) predicts a positive correction that vanishes when $T$ is large: on the contrary we observe a convergence to a value lower than $H_1^{\rm fBm}$ which hints that the bias is negative and larger than the theoretical correction.
This suggests that the form of eq.(\ref{eq:h_T:fBm}) is still valid for embedding dimensions $m>1$.

\subsubsection{Entropy rate dependence on scale $\tau$}

Fig.~\ref{fig:h:tau:fBm}b shows that for a fixed window size $T=2^{16}$ the ersatz entropy rate is proportional to $\mathcal{H}\ln(\tau)$. We have added a black line defined by the linear function $H_1^{\rm FBM} + \mathcal{H} \ln \tau$, as suggested by eq.(\ref{eq:h_T:fBm}) without the corrective term. This black line perfectly describes the evolution of the entropy rate with the scale $\tau$, which is independent on the embedding dimension $m$.

To observe the finer evolution of the entropy rate on the scale $\tau$, we subtract the main contribution $\mathcal{H} \ln \tau$ to the entropy rate and we plot $\bar{h}_{T}^{(m,\tau)} - \mathcal{H} \ln \tau$ for different embedding dimensions in Fig.~\ref{fig:h-hurst}.
We observe a slight increase, which is larger for larger embedding dimensions.
For $m=1$, the correction term can be evaluated from eq.(\ref{eq:correction:full}), and is at most $2.10^{-3}$, and does not account for the evolution reported here, which is probably due to the bias which increases when the number of points ---~which is proportional to $T/\tau$~--- decreases and when the embedding $m$ increases.

For a scale invariant self similar process, the standard deviation $\sigma_\tau$ of the increments of size $\tau$ behaves as $\sigma_\tau = \sigma_1 \tau^{\cal H}$.
Subtracting $\mathcal{H} \ln \tau$ amounts to subtracting $\ln \sigma_\tau$: for each scale $\tau$, this corresponds to normalizing the down-sampled data (taking one point every $\tau$ points) by the standard deviation $\sigma_\tau$ of the increments of size $\tau$.
When the Hurst exponent is {\em a priori} unknown, $\sigma_\tau$ can be computed, and used to compute the main contribution $-\ln(\sigma_{\tau})$; thus the fine evolution of the entropy rate with $\tau$ can be used as a tool to probe the deviation from the self similarity assumption, which is interesting for multifractal signals.

%%%%%%%%%%%%%%%%%%%%%%%%%%%%%%%%%%%%%%%%%%%
\section{Application of the practical framework to non-Gaussian self-similar processes}
\label{sec:application}
\label{sec:log-normal}

In this section, we turn to non-Gaussian processes and describe the results obtained in the time-average framework generalized in this larger class of processes.

\paragraph{Procedure}

We construct two different motions, in the very same way as we did for the fBm. We integrated two log-normal noises synthesized with the same log-normal marginal distribution and with the same correlation function (\ref{eq:fGn:correlation}) as the fGn, but different dependance structure. 
To generate these noises,  we use the methodology proposed in \cite{Helgason2011} to obtain the log-normal marginal by applying two different transformations to the cumulative distribution function $F_Z$ of a Gaussian white noise $Z$: the Hermitian transformation of rank 1 ($f^{1}(z)=F^{-1}(F_Z(z))$) and the even-Hermitian transformation of rank 2 : $f^{1}(z)=F^{-1}(2(F_Z(|z|)-\frac{1}{2}))$, where $F$ is the cumulative distribution function of the targeted log-normal distribution.
This synthesis is performed with the toolbox provided at \href{http://www.hermir.org}{www.hermir.org}. 
Once the two log-normal noises have been generated, they are integrated using eq.(\ref{eq:motion}) to obtain two non-stationary scale invariant processes with non-Gaussian statistics.

The dependence structures of the two log-normal noises were previously studied in detail~\cite{Granero-Belinchon2019}: while the correlation function is the same for the two noises ---~and identical to the targeted one of the fBm given by (\ref{eq:fGn:correlation})~--- the complete dependence structure was shown to be different.

To study these two non-stationary and non-Gaussian motions, we use again realizations of $T=2^{16}$ points, $k=5$ neighbors and we focus on the case where embedding dimension $m=1$ and Hurst exponent ${\cal H}=0.7$.

%%%%
\paragraph{Bias and standard deviation}

We report in Fig.~\ref{fig:bias}b and \ref{fig:bias}c the evolution of the ersatz entropy rate of the Hermitian and the even-Hermitian log-normal processes in function of $\frac{k}{T^{1/2}}$.
We observe exactly the same behavior as for the fBm: the entropy rate converges to $H_1^{\rm ln}$, the entropy of the log-normal process at unit-time\footnote{if $X$ is a log-normal process of mean $\mu$ and standard deviation $\sigma$, then the process $\log X$ is Gaussian with the mean $\mu'=\log\left(\frac{\mu^2}{\sqrt{\mu^2+\sigma^2}}\right)$ and the standard deviation $\sigma'=2\log\left(1+\frac{\sigma^2}{\mu^2}\right)$ and the entropy of $X$ can be expressed as~\cite{Granero-Belinchon2019}: $$H_1^{\rm ln} = \frac{1}{2}\log(2\pi e \sigma') + \mu'\,.$$} (horizontal blue/red line in Fig.~\ref{fig:bias}b,c), which then gives an estimation of the bias of our estimator, which appears to be the same as for the fBm.
 
We report in Fig.~\ref{fig:std}b and Fig.~\ref{fig:std}c the behavior of the standard deviation of the estimators. Again, exactly as for the fBm, the standard deviation is large for the ersatz entropy and the ersatz auto-mutual information, while it is much smaller for the ersatz entropy rate.

Again, both the bias and the standard deviation of the entropy rate increase when $k$ increases or $T$ decreases and can be made arbitrarily small by increasing $T$. These results do not depend on the marginal distribution: they have been obtained not only for the fBm with Gaussian statistics, but also for two motions built on log-normal noises.
 
\paragraph{Dependence on times $T$ and $\tau$} 

\begin{figure}
\begin{center}
\includegraphics[width=\linewidth]{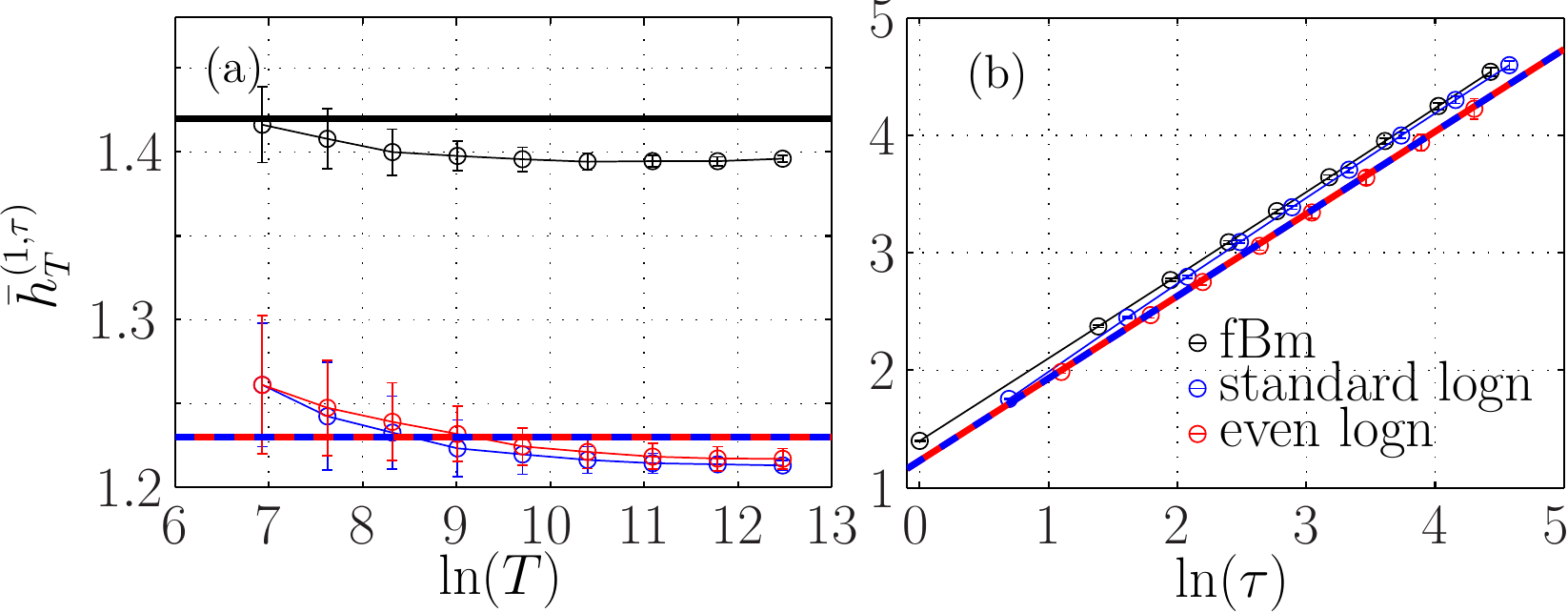}
\caption{$\bar{h}_{T}^{(1,\tau)}$ for a motion built from a Hermitian (blue) or even-Hermitian (red) log-normal noise, as a function of a) the time window size $T$ or b) the time scale $\tau$.
Results for the fBm (from Fig.~\ref{fig:h:tau:fBm} with $m=1$) are reported in black for comparison.
$T=2^{16}$ and $k=5$. 
The horizontal lines in (a) indicates the entropy $H_1$ of the noise (in black for the fBm and in red and blue for a log-normal process).
}
\label{fig:lognormal:h}
\end{center}
\end{figure}

The evolution of $\bar{h}_{T}^{(1,\tau)}$ on the time window size $T$ for the two motions is presented in Fig.~\ref{fig:lognormal:h}a). As it was the case for the fBm, $\bar{h}_{T}^{(1,\tau)}$ depends only weakly on $T$, and seems to converge for larger $T$ to the value $H_1^{\rm ln}$, up to a small corrective term.

\begin{figure}
\begin{center}
\includegraphics[width=\linewidth]{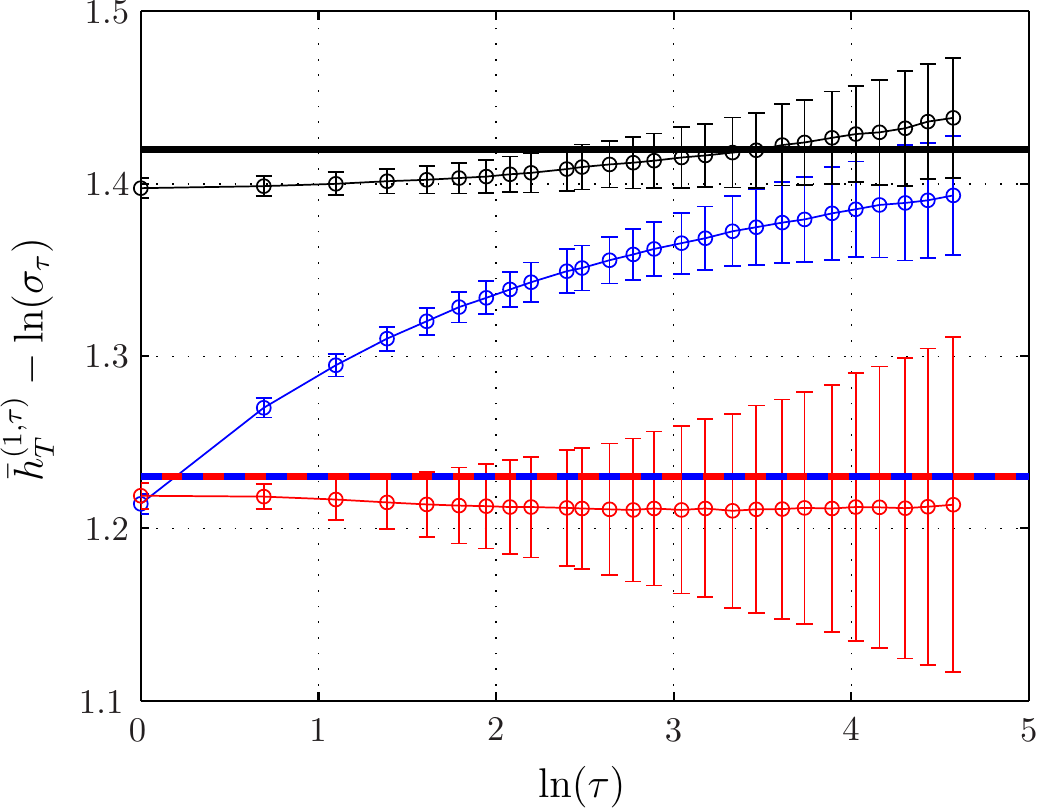}
\caption{Ersatz entropy rate $\bar{h}_{T}^{(m=1,\tau)}-\ln(\sigma_{\tau})$ for motions built on Hermitian (blue) or even-Hermitian (red) log-normal noise, together with results for the fBm (black) as a function of $\ln\tau$. 
$T=2^{16}$ and $k=5$. 
The straight lines in (b) indicate the theoretical values of the entropy of the processes. 
}
\label{fig:lognormal:hnormalized}
\end{center}
\end{figure}

The evolution of $\bar{h}_{T}^{(1,\tau)}$ with the time scale $\tau$ is presented in Fig.~\ref{fig:lognormal:h}b). 
In the same way as for the fBm, we again observe a large increase, almost proportional to $\ln\tau$. Because this strong tendency originates from the increase of the standard deviation $\sigma_\tau$ of the increments of size $\tau$ when $\tau$ increases, we again normalize the entropy rate by subtracting $\ln\sigma_\tau={\cal H}\ln\tau$. Results are presented in Fig.~\ref{fig:lognormal:hnormalized}, together with results for the fBm with $m=1$ for comparison.

The normalized ersatz entropy rate of the motion built from the even-Hermitian log-normal noise appears as almost independent of $\tau$. This behavior is identical to the one observed for the fBm, but the remaining constant value is different ($H_1^{\rm fBm}$ or $H_1^{\rm ln}$).
%Their values at $\tau=1$ are directly related to the probability density function of the noises defining the motions, see equation~\ref{eq:hxHdx} for the \textit{general framework}
The ersatz entropy rate of the fBm (in black) and the even-Hermitian motion (in red) both behaves exactly as ${\cal H}\ln\tau$, which is the expected behavior for a self-similar process, see eq.(\ref{eq:h:abstract:ssimilar}). On the contrary, the motion built with the Hermitian transformation of rank 1 exhibits an additional variation in $\tau$: the normalized entropy rate $\bar{h}_{T}^{(m,\tau)}-\ln(\sigma_{\tau})$ evolves from the value $H_1^{\rm ln}$ at $\tau=1$ ---~expected for the motion built with a log-normal noise and obtained for the even-Hermitian process at any $\tau$~--- up to the value $H_1^{\rm fBm}$ ---~expected for a Gaussian process, and obtained for the fBm at any $\tau$.

\begin{figure}[t]
\begin{center}
\includegraphics[width=\linewidth]{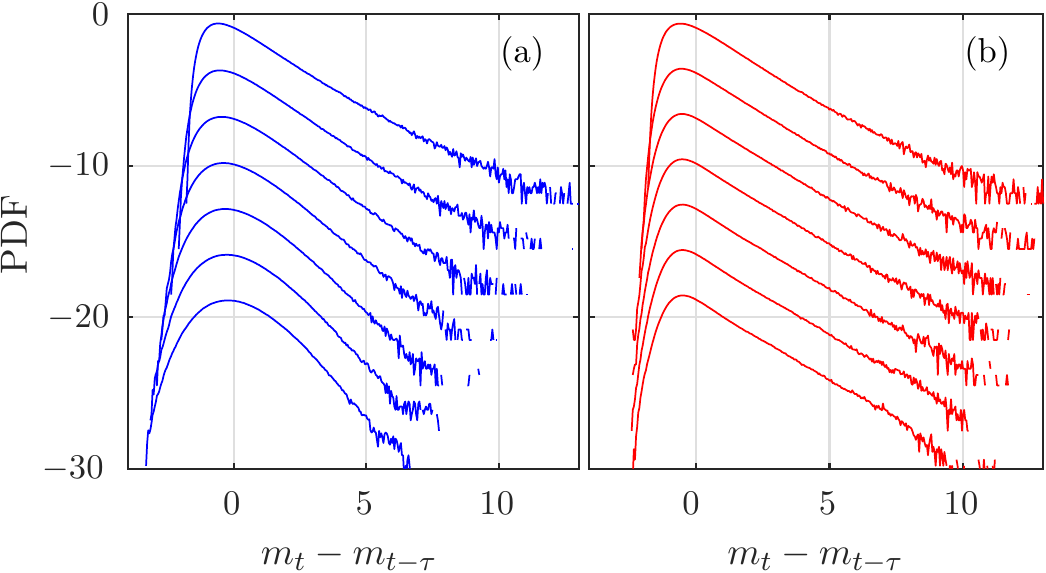}
\caption{PDF of the increments of the (a) Hermitian and (b) even-Hermitian log-normal motions of size $\tau=2^j$, from $j=0$ (bottom) up to $j=6$ (up). Curves have been arbitrarily shifted on the Y-axis for clarity.}
\label{fig:lognormal:histograms}
\end{center}
\end{figure}

As a conclusion, one can estimate the Hurst exponent of a perfectly self-similar process as the slope of the linear fit in $\ln\tau$ of the ersatz entropy rate. This is a valid approach for the fBm and the motion built from the noise constructed with the even-Hermitian transformation, because the ersatz entropy rate then behaves linearly in $\ln \tau$.
On the contrary, the motion built using an hermitian transformation of rank 1 does not appear as perfectly self-similar. This can be indeed verified by plotting the normalized PDFs (setting the standard deviation to unity) of the increments $m_t - m_{t-\tau}$ of the motions for various values of $\tau$.
As can be seen in Fig.~\ref{fig:lognormal:histograms} the PDFs of the increments of the "standard log-normal process" varies with the scale $\tau$, while these of the "even-Hermitian motion" remain identical. For $\tau=1$, the increments are nothing but the log-normal noises, which are log-normal, as prescribed. For large $\tau$, the increments of the "even-Hermitian motion" remain log-normal, while the increments of the standard log-normal motion" deforms and seems to become more Gaussian. The ersatz entropy rate catches this fine evolution perfectly.

%%%%%%%%%%%%%%%%%%%%%%%%%%%%%%%%%%%%%%%%%%%%%%%%%%%%%%%%%%%%%%%%%
\section{Application of the practical framework to a multifractal process}
\label{sec:multifractal}

We now explore the proposed time-averaged framework on the Multifractal Random Walk, to illustrate how it performs on a multifractal process.
The multifractal random walk (MRW)~\cite{Bacry2001,Bacry2002} is a popular multiplicative cascade process widely used to model systems that exhibit multifractal properties~\cite{Delour2001}. 
Like the fBm, the MRW is a motion obtained by integrating ---~again with eq.~(\ref{eq:motion})~--- a stationary noise $W^{\rm MRW}=\{w_t^{\rm MRW}\}_{t\in \R}$ such that 
\begin{equation}
w_t^{\rm MRW} = w^{\rm fGn}_t e^{\omega_t}
\end{equation}
where $W^{\rm fGn}=\{w^{\rm fGn}_t\}_{t\in \R}$ is a fGn with parameter ${\cal H}^{\rm fGn}$ 
and $\Omega=\{\omega_t\}_{t\in \R}$ is a Gaussian random process, independent of  $X^{\rm fGn}$ with a correlation function 
\begin{align}
c_\omega(\tau) &= -c_2 \log\left(\frac{L}{|\tau|+1}\right) &{\rm if} \quad |\tau|<L \\
 &=0 & {\rm otherwise} \nonumber
 \end{align}
where $L$ is the integral scale, set here to $L=T$.

The MRW is a scale invariant process: 
the power spectrum of its time-derivative $W^{\rm MRW}$ behaves as a power law with an exponent $2({\cal H}^{\rm fGn}-c_2)+1$, which would be the Hurst exponent obtained for a fGn with parameter ${\cal H}^{\rm fGn}-c_2$. 
Any moment of order $q$ of the increments of size $\tau$ behaves as a power law of $\tau$ with the exponent $\zeta(q)$.
Contrary to the fBm, the MRW is not exactly self similar and exhibits intermittency: $\zeta(q)={\cal H}^{\rm fGn}q - \frac{c_2}{2}q^2$ is not a linear function of $q$, as expected for a self-similar process. 
As a consequence, the shape of the PDF of the increments depends on the scale.  

\begin{figure}[t]
\begin{center}
\includegraphics[width=\linewidth]{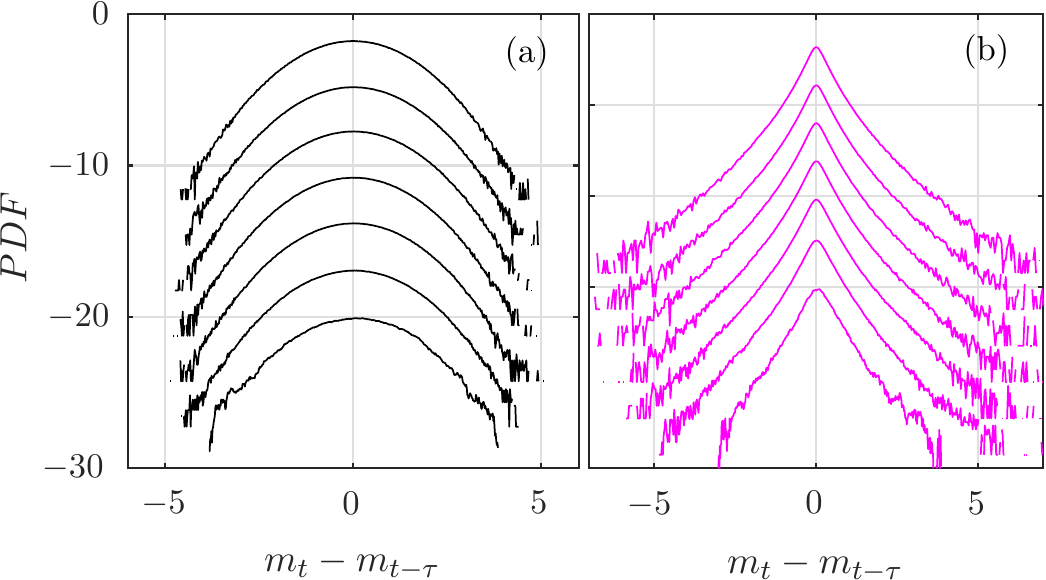}
\caption{PDF of  the increments of the (a) fBm and (b) of a MRW  of size $\tau=2^j$, from $j=0$ (bottom) up to $j=6$ (up). Curves have been arbitrarily shifted on the Y-axis for clarity.}
\label{fig:MRW:histograms}
\end{center}
\end{figure}

We choose the parameter ${\cal H}^{\rm fGn}$ such that the power spectrum of the noise $W^{\rm MRW}$ is identical to the one of the fBm used in the former sections, {\em i.e.}, ${\cal H}^{\rm fGn}=0.7+c_2$.
We set the parameter $c_2=0.025$, a value widely used to model the intermittency of Eulerian turbulent velocity field~\cite{Chevillard2012}.

Figure \ref{fig:MRW:histograms} compares the evolution of the PDF of the increments of the fBm and the MRW.
As expected, no change is observed for the fBm, while the PDF of the MRW has wider tails for smaller $\tau$. 
The fBm is perfectly self-similar, while the MRW exhibits intermittency~\cite{Granero-Belinchon2018}: the PDF of its increments is deformed when the scale $\tau$ of the increments is varied, although no analytical expression of the PDF is available.

\begin{figure}
\begin{center}
\includegraphics[width=\linewidth]{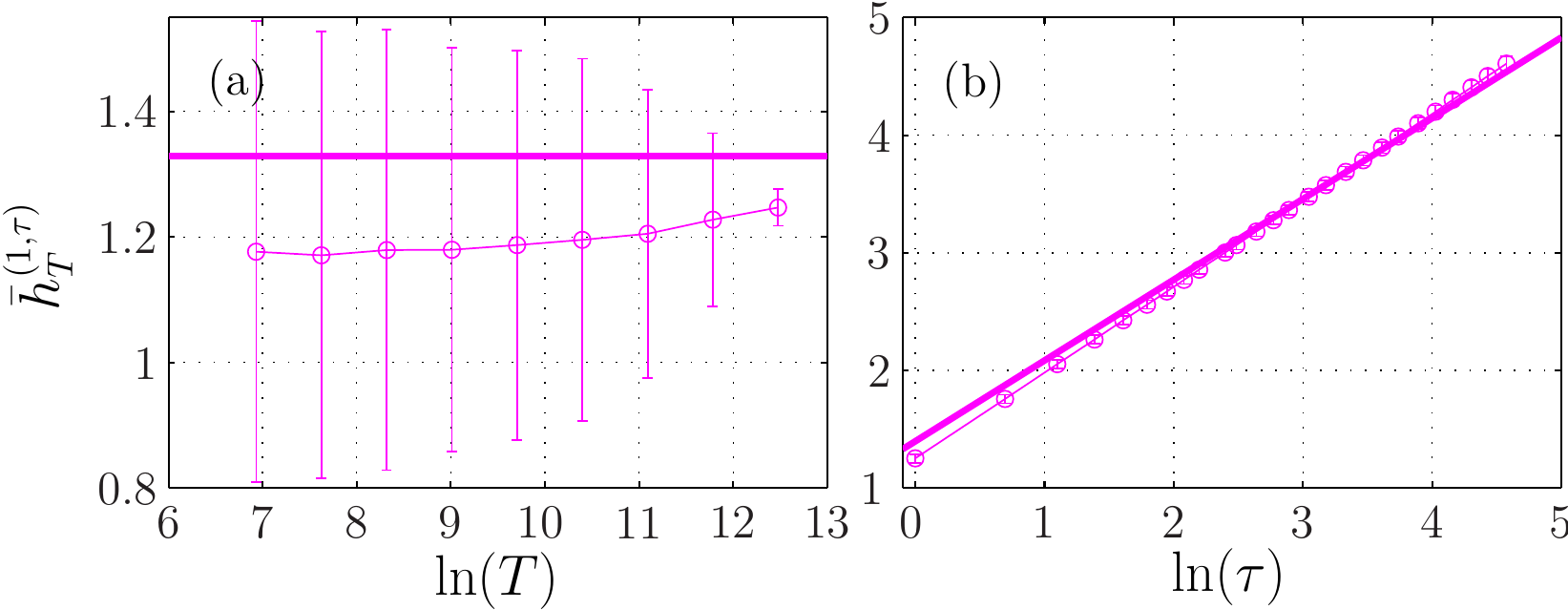}
\caption{Ersatz entropy rate $\bar{h}_{T}^{(m=1,\tau)}$ of a MRW with $\mathcal{H}=0.7$.
(a): as a function of the window size $T$ for fixed $\tau=1$.
(b): as a function of the scale $\tau$ for $T=2^{16}$. 
%In (a) and (c), the entropy rate is computed using eq.(\ref{eq:hT}), while in (b) and (d), it is computed using eq.(\ref{eq:hT2}).
The horizontal line in (a) indicates the numerical value $H_1^{\rm MRW}$ of the noise.
The straight line in (b) has a slope $\mathcal{H}=0.7$.
}
\label{fig:MRW:h}
\end{center}
\end{figure}

We apply our practical framework  and plot in Fig.~\ref{fig:MRW:h}a) the evolution of the ersatz entropy rate of the MRW with $T$. 
Again, the entropy rate seems to be independent of $T$. We nevertheless observe a small tendency to increase towards the value $H_1^{\rm MRW}$ the entropy of the MRW at unit-time. Here, because there is no analytical expression of the PDF, $H_1^{\rm MRW}$ cannot be derived analytically and we numericaly estimate its value.
% the convergence is slower, and may not be due only to the estimator, but also to the complex dynamics of the process
 
\begin{figure}
\begin{center}
\includegraphics[width=\linewidth]{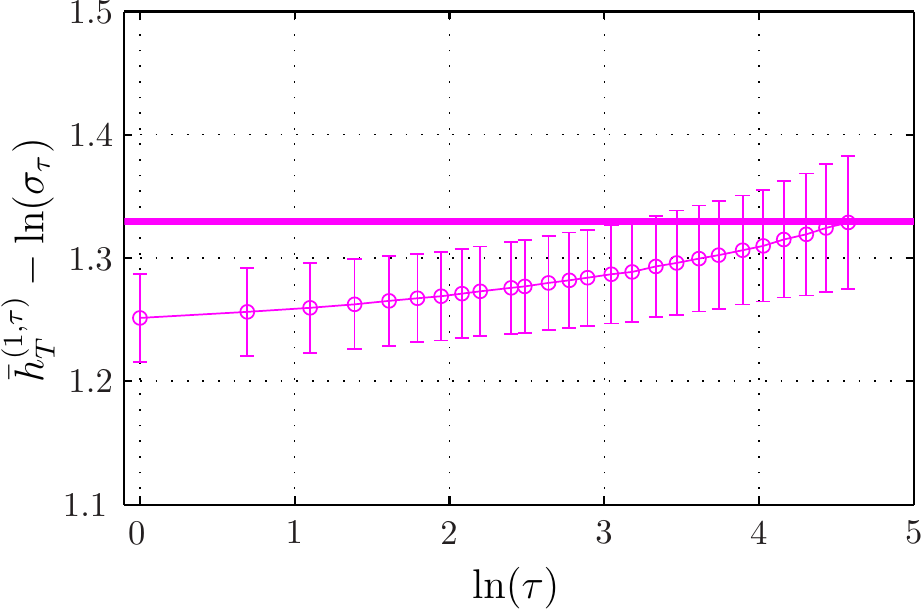}
\caption{Ersatz entropy rate $\bar{h}_{T}^{(m=1,\tau)}-\ln(\sigma_{\tau})$ of the MRW as a function of $\ln(\tau)$ for fixed $T=2^{16}$.
}
\label{fig:MRW:hnormalized}
\end{center}
\end{figure}

The dependence in $\tau$ is plotted in Fig.~\ref{fig:MRW:h}b). We again observe a strong linear evolution of the ersatz entropy rate in ${\cal H}\ln\tau$. After subtracting this strong tendency (Fig.~\ref{fig:MRW:hnormalized}), we still observe an evolution with $\tau$, but this evolution appears much weaker than for the Hermitian log-normal (blue curve in Fig.~\ref{fig:lognormal:hnormalized}). Indeed, the deformation of the PDFs of the increments when varying $\tau$ is much slower for the MRW (Fig.~\ref{fig:MRW:histograms}b) than for the Hermitian log-normal (Fig.~\ref{fig:lognormal:histograms}a).

%%%%%%%%%%%%%%%%%%%%%%%%%%%%%%%%%%%%%%%%%%%%%%%%%%%
\section{Discussion and Conclusions}
\label{sec:discussion}

We proposed a new framework in information theory to analyze a non-stationary process by considering it as resulting from a gedanken stationary process and estimating the PDF by cumulating all available samples in a time interval of size $T$. This framework hence considers a PDF obtained by time-averaging over a time window $[t_0: t_0+T]$, and then proceeds to compute the associated information theory quantities.
In particular, the ersatz entropy $\bar{H}_{T}(X)$ that is then defined can be interpreted as the amount of information characterizing the complete trajectory $\{X_t, t \in [t_0,t_0+T]\}$ of the process $X$. 
If we assume that the increments of $X$ are stationary and centered, then $\bar{H}_{T}(X)$ and all other information ersatz theoretical quantities depend only on the duration $T$ and not of the first time $t_0$.

We illustrated our approach by focusing first on a model system: the fractional Brownian motion. We derived in this context the analytical expressions of the ersatz entropy, ersatz auto-mutual information and ersatz entropy rate, which allowed a pedagogical description of our new information theory quantities. 
We also reported how the ersatz quantities behave when the time-interval size $T$ and the embedding time scale $\tau$ are varied: we obtained analytical expressions for embedding dimension $m=1$, and confirmed them numerically for $m\ge 1$. Besides the fBm, we reported numerical observations for various self similar or multifractal processes. The ersatz entropy $\bar{H}_{T}^{(m,\tau)}$ always diverges logarithmically in $T$ while the ersatz entropy rate $\bar{h}_{T}^{(m,\tau)}$ always behaves as almost independent of $T$. The examination of how the ersatz entropy rate $\bar{h}_{T}^{(m,\tau)}$ depends on the scale $\tau$ provides a fine exploration of either the self-similarity or the multifractality of the process.

This exploration of the multifractality of a non-stationary process with stationary increments using the ersatz entropy rate $\bar{h}_{T}^{(m,\tau)}(M)$ gives a viewpoint very similar to the one reported when analyzing the increments of the process with the regular Shanon entropy, as reported in \cite{Granero-Belinchon2018}. We are currently investigating how to relate quantitatively the two approaches.

In the same vein, the ersatz entropy rate allowed us to discriminate two different non-stationary processes, and obtain fine differences in their self similarity properties (figure~\ref{fig:lognormal:hnormalized}), in close relation to a method using the entropy rate of the increments of the signal, as exposed in \cite{Granero-Belinchon2019}. A possible connection is also under investigation.

Through all this article, we have estimated the ersatz quantities of a process on a single trajectory $[t_0; t_0+T]$ of this process; this situation corresponds to the worst case scenario where only a single realization of the process is know. If enough experimental data are available, one can improve the estimation of the ersatz quantities in two ways. First, if the same experiment has been conducted multiple times, and thus multiple realizations are available over the time interval $[t_0; t_0+T]$, one can use all these independent realizations to enhance the estimation of the time-averaged PDF. Second, if a single but long enough realization of size ${\cal T}$ is available, one can split it into multiple time intervals $[kT; (k+1)T]$,  $k \in [0 .. \lfloor{\cal T}/T\rfloor]$ and the use these intervals as independent realizations as in the first case. This later situation is made possible by the assumption that the increments of the signal are not only stationary, but also centered.

\appendix
\section{Entropy of a time-embedded signal}
\label{sec:appendix}
%This follows from expressing the movement $M$ as a sum of its "recent" increments, and using chained conditioned probabilities (XII.164).

The time-embedded vector $\textbf{x}_t^{(m,\tau)}$ (eq.~\ref{eq:embed}) can be mapped into the vector
$\tilde{\textbf{x}}_t^{(m,\tau)}\equiv (x_t, \delta_\tau x_t, \delta_\tau x_{t-\tau}, ..., \delta_\tau x_{t-(m-2)\tau} )$ by the linear transformation $Q^m$:
\begin{align}
\textbf{x}_t^{(m,\tau)} &\mapsto \tilde{\textbf{x}}_t^{(m,\tau)} = Q^m. \textbf{x}_t^{(m,\tau)} ,
\end{align}
where $Q^m$ is the band matrix defined as:
\begin{align}
Q^m \equiv \left(\begin{matrix}
1 &  0 & 0 & \cdots & 0 & 0 \\
1 & -1 & 0 & \cdots & 0 & 0 \\
0 & 1 & -1 & \cdots & 0 & 0 \\
\vdots & \vdots & \vdots & \ddots & \vdots & \vdots \\
0 & 0 & 0 &  \cdots & -1 & 0\\
0 & 0 & 0 &  \cdots & 1 & -1 \\
\end{matrix}\right) \,.
\label{def:matrix}
\end{align}

which determinant satisfies $|{\rm det}(Q_m)|=1$. As a consequence, $H(\tilde{\textbf{x}}_t^{(m,\tau)}) = H(\textbf{x}_t^{(m,\tau)})$, which proves (\ref{eq:H:inc}).

\acknowledgments{
The authors wish to thank L. Chevillard for stimulating discussions.
This work was supported by the LABEX iMUST (ANR-10-LABX-0064) of Universit\'e de Lyon, within the program "Investissements d'Avenir" (ANR-11-IDEX-0007) operated by the French National Research Agency (ANR).
}

\reftitle{References}
\bibstyle{mdpi}
\bibliography{THEBIBLIO}
%\bibliography{biblio-IEEE}

\end{document}